\documentclass[showpacs,preprintnumbers,amsmath,amssymb]{revtex4}

\usepackage{amsmath}
\usepackage[dvips]{graphicx}

\makeatletter
\renewcommand{\fnum@table}{\textbf{\tablename~\thetable}}
\renewcommand{\fnum@figure}{\textbf{\figurename~\thefigure}}
\makeatother

\newcommand {\be}{\begin{equation}}
\newcommand {\ee}{\end{equation}}
\newcommand {\ba}{\begin{eqnarray}}
\newcommand {\ea}{\end{eqnarray}}

\begin{document}


\title{}
\vspace*{10mm}
\title{Minimal model linking two great mysteries:
Neutrino mass and dark matter}
\author{Yasaman Farzan}\email{yasaman@theory.ipm.ac.ir}
\affiliation{Institute for research in fundamental sciences (IPM),
P.O. Box 19395-5531, Tehran, Iran}

\begin{abstract}
We present an economic model that establishes a link between
neutrino masses  and  properties of the dark matter candidate. The
particle content of the model can be divided into two groups:
light particles with masses lighter than the electroweak scale and
heavy particles. The light particles, which also include the dark
matter candidate, are predicted to show up in the low energy
experiments such as $(K\to \ell +{\rm missing~energy})$, making
the model testable. The heavy sector can show up at the LHC and
may give rise to Br($\ell_i \to \ell_j \gamma$) close to the
present bounds. In principle, the new couplings of the model can
independently be derived from the data from the LHC and from the
information on neutrino masses and Lepton Flavor Violating (LFV)
rare decays, providing the possibility of an intensive cross-check
of the model.
\end{abstract}
\date{\today}
\maketitle

\section{Introduction}

The nature of dark matter and the tiny neutrino masses are two
great mysteries of modern particle physics and cosmology. A myriad
of models have been proposed in the literature to explain either
of these phenomena. Recently there have been some attempts to link
these two phenomena \cite{celine,linkingModels}.

 In Ref.~\cite{celine}, a scenario
has been proposed  that  most economically  accounts  for the
neutrino mass and simultaneously  provides a dark matter
candidate. The new particle content of the scenario is composed of
a scalar field $\phi$ which plays the role of dark matter and two
(or more) right-handed Majorana neutrinos, $N_{i }$. A $Z_2$
symmetry has been introduced under which $\phi$ and $N_{i}$ are
odd but the standard model particles are even. The $Z_2$ symmetry
stabilizes the lightest new particle, making it a suitable dark
matter candidate. Moreover, the $Z_2$ symmetry forbids the
$\epsilon_{\alpha \beta} \bar{N} H^T_\alpha L_\beta$ term in the
Lagrangian so the neutrinos do not obtain a Dirac mass term.

Following \cite{celine}, we shall call the scalar a {\it SLIM}.
The scenario can be realized in two cases: i) real scalar; ii)
complex scalar.  In the present paper, we shall focus on the
scenario with real SLIM.
 The
scenario is based on the following low energy  effective
Lagrangian: \be \label{Lcal} \mathcal{L}=-g_{i \alpha} \bar{N}_{i
} \nu_{\alpha L} \phi- \bar{N}_{j}^c {(M_N)_{ij} \over 2} N_{i}\
.\ee This Lagrangian at one loop level gives Majorana masses to
the active neutrinos: $m_\nu=(g^2M_N/16\pi^2)\log
\Lambda^2/M_N^2$. The dominant annihilation modes of the dark
matter is into a neutrino or antineutrino pair:
$\sigma_{tot}\simeq \sigma(\phi \phi \to \nu \nu)+\sigma(\phi \phi
\to \bar{\nu} \bar{\nu})$ . Under the assumption that the
production of dark matter in the early universe was thermal, the
dark matter abundance determines the annihilation cross section
($\langle \sigma_{tot} v_r\rangle \sim 3 \cdot 10^{-26}~{\rm
cm^3/sec}$), which is in turn   given by the coupling $g$ and the
masses of the right-handed neutrinos. The same parameters also
determine the Majorana masses of active neutrinos:
$\sigma_{tot}=(g^4/2\pi)[m_N^2/(m_\phi^2+m_N^2)^2]$. Combining
these two pieces of information, the parameters of the model can
be restricted.  In case of real scalar, the mass of the lightest
right-handed neutrino is found to be in the range of few to
10~MeV. Remember that the scalar  is taken to be even lighter than
the right-handed neutrinos. On the other hand, a lower bound of
few$\times 10^{-4}$ is derived on the coupling \cite{celine}. With
the lower bound on the coupling and the upper bound on the masses
of new particles, the scenario can be tested with experiments and
observations that are sensitive to rare but low energy processes;
{\it e.g.}, studies of supernova core collapse and searches for
rare decays of light mesons such as the Kaon or pion
\cite{celine}.

 The effective Lagrangian (\ref{Lcal}) is valid only
 in low energies and has to be
embedded within a high energy theory which is invariant under
SU(3)$\times$SU(2)$_L\times$U(1). Recently, a model has been
proposed that embeds the scenario with a complex SLIM
\cite{ourselves}. In \cite{Ma}, a model has been proposed that can
be considered a realization of complex SLIM but the question of
dark matter abundance has not been addressed in \cite{Ma}. In the
present paper, we propose a minimalistic model which embeds the
scenario with a real SLIM.

The paper is organized as follows. In sect.~\ref{RealSLIM}, we
introduce the model. In sect.~\ref{Impli}, we discuss the
phenomenological implications of the model. The results are
summarized in sect.~\ref{con}.
\section{Real SLIM \label{RealSLIM}}
The content of this model is composed of (1) an electroweak
singlet, $\eta$; (2) two Majorana right-handed  neutrinos, $N_{i}$
and (3) an electroweak doublet with nonzero hypercharge,
$\Phi^T=[\phi^0 \ \phi^-]$ where $\phi^0\equiv
(\phi_1+i\phi_2)/\sqrt{2}$ with real $\phi_1$ and $\phi_2$.
Similarly to the scenario in \cite{celine}, all these particles
are odd under the $Z_2$ symmetry. The most general $Z_2$ even
renormalizable Lagrangian involving only the scalars can be
written as
\begin{align}
\mathcal{L}=&-m_\Phi^2 \Phi^\dagger \cdot \Phi -{m_s^2\over 2}
\eta^2
- (m_{\eta \Phi} \eta(H^T (i\sigma_2) \Phi)+{\rm H.c.}) \nonumber \\
~&-{\lambda_1 } |H^T (i\sigma_2) \Phi|^2-{\rm Re}[\lambda_2 (H^T
(i\sigma_2) \Phi)^2]-\lambda_3 \eta^2 H^\dagger H-\lambda_4
\Phi^\dagger \cdot \Phi H^\dagger \cdot H
\nonumber \\
~& -{\lambda'_1 \over 2} (\Phi^\dagger \cdot
\Phi)^2-{\lambda'_2\over 2} \eta^4-\lambda'_3\eta^2 \Phi^\dagger
\cdot \Phi \nonumber \\
~& -m_H^2 H^\dagger\cdot H- {\lambda \over 2}(H^\dagger \cdot H)^2
\label{MainLag}
\end{align}
Certain conditions on the parameters of the model should be
satisfied so that the theory becomes stable in the sense that when
a combination of the fields goes to infinity, the potential
remains positive \cite{G-H}. Some of these conditions are
$$\lambda'_1,\lambda'_2>0, \
\lambda'_3>-(\lambda'_1\lambda'_2)^{1/2}, \
\lambda_3>-(\lambda\lambda'_2)^{1/2}$$ and $$
\lambda_1-|\lambda_2|+\lambda_4 > -(\lambda\lambda'_1)^{1/2}\ . $$

 For simplicity
we take the Lagrangian to be CP-conserving which means $m_{\eta
\Phi}$ and $\lambda_2$ are both real. As we shall see below, after
electroweak symmetry breaking, the term $m_{\eta \Phi}$ mixes
$\eta$ with the CP-even neutral component of $\Phi$ ({\it i.e.,}
$\phi_1$). The lightest new particle is a linear combination of
$\eta$ and $\phi_1$ with a dominant contribution from $\eta$.
Because of the $Z_2$ symmetry, such a combination is stable and
plays the role of the dark matter [{\it i.e.,} the role of $\phi$
in Ref. \cite{celine}; see Eq.~(\ref{Lcal})].

The Lagrangian involving the right-handed neutrinos in the mass
basis of right-handed neutrinos is \be \mathcal{L}=-g_{i
\alpha}\bar{N}_{i} \Phi^\dagger\cdot L_\alpha -{M_{i}\over 2}
\bar{N}_i^c N_i\ \label{sterileL},\ee where  $L_\alpha$ is the
lepton doublet of flavor $\alpha$: $L_\alpha^T=(\nu_{L\alpha} \
\ell_{L\alpha}^-)$.

 After electroweak symmetry breaking ({\it i.e.,}
setting $H^T=(0 \ v_H/\sqrt{2})$), the mass terms will be of form
\begin{align}
\mathcal{L}_m&=-m_{\phi^-}^2|\phi^-|^2- \frac{m_{\phi_2}^2}{2}\phi_2^2\\
&-\frac{m_\eta^2}{2}\eta^2-\frac{m_{\phi_1}^2}{2} \phi_1^2
-m_{\eta \Phi}v_H \phi_1 \eta
\end{align} where
\begin{align}
m_{\phi^-}^2=&
m_\Phi^2+\lambda_4\frac{v_H^2}{2}\label{phi-}\\
m_\eta^2=&m_s^2+\lambda_3\frac{v_H^2}{2}\\
m_{\phi_1}^2 =& m_\Phi^2+\lambda_1\frac{v_H^2}{2}+
\lambda_2\frac{v_H^2}{2} \\
m_{\phi_2}^2
=&m_\Phi^2+\lambda_1\frac{v_H^2}{2}-\lambda_2\frac{v_H^2}{2}\ .
\label{phi2}
\end{align}
The parameters can be tuned such that the squares of all mass
eigenvalues become positive. This means that other than the Higgs
field, none of the scalars develops a vacuum expectation value  so
the $Z_2$ symmetry remains unbroken and moreover neutrinos do not
acquire any Dirac mass at the tree level. $\phi^-$ and $\phi_2$
are both mass eigenvalues whose masses can be readily read
respectively from Eq.~(\ref{phi-}) and Eq.~(\ref{phi2}). To avoid
the bounds from direct searches, $m_{\phi^-}^2$ and $m_{\phi_2}^2$
are taken above the electroweak scale. Assuming the couplings of
Higgs are relatively small ({\it i.e.,} $\stackrel{<}{\sim} 0.1$),
this means $m_\Phi^2$, and consequently $m_{\phi_2}^2$,
$m_{\phi_1}^2$ and $m_{\phi^-}^2$, are of order of (or larger
than) $O((100~{\rm GeV})^2)$. However, $m_\eta^2$ can be below the
electroweak scale.
  The mass eigenstates are \ba \label{mixing-ETA-PHI1} \left[
\begin{matrix}\delta_1 \cr \delta_2 \end{matrix}\right]=
\left[
\begin{matrix} \cos \alpha & -\sin \alpha \cr
\sin \alpha & \cos \alpha \end{matrix} \right]\left[
\begin{matrix} \eta \cr \phi_1 \end{matrix} \right]\ea with
\begin{align}
\tan 2 \alpha &= {2 v_H m_{\eta \Phi} \over
m_{\phi_1}^2-m_\eta^2}\label{2ALPHA}\\ m^2_{\delta_1} &\simeq
m_\eta^2- {(m_{\eta
\Phi} v_H)^2 \over m_{\phi_1}^2-m_\eta^2}\\
m^2_{\delta_2} &\simeq m_{\phi_1}^2+ {(m_{\eta \Phi} v_H)^2 \over
m_{\phi_1}^2-m_\eta^2}\ ,\end{align} where in the last two
equations we have used $(m_{\eta\Phi}
v_H)^2/(m_{\phi_1}^2-m_\eta^2)^2\ll 1$. The couplings of the mass
eigenvalues $\delta_1$ and $\delta_2$ to  $\bar{N}_i\nu_{L\alpha}$
are listed in Table \ref{tab:partcont}. We are interested in the
following range: \be \label{con1} m_{\delta_1}^2 <m_{N_1}^2\ll
m_{\delta_2}^2\simeq m_{\phi_2}^2 \simeq m_{\phi^-}^2 \sim
m_{electroweak}^2  \ee {\rm and} \be \label{con2}\left|
{m_{\phi_2}^2-m_{\delta_2}^2 \over
m_{\phi_2}^2+m_{\delta_2}^2}\right| \simeq \left|-{\lambda_2 \over
2}{v_H^2 \over m_{\phi_2}^2} -{\sin^2 \alpha \over 2}\right| \ll 1
\ . \ee $\delta_1$, being the lightest $Z_2$-odd particle, is the
dark matter candidate. Using the couplings in
Table~\ref{tab:partcont}, we find \be \langle
\sigma(\delta_1\delta_1 \to \nu_{L\alpha} \nu_{L \beta})
v_r\rangle =\langle \sigma(\delta_1\delta_1 \to \bar\nu_{L\alpha}
\bar\nu_{L \beta}) v_r\rangle= {\sin^4 \alpha \over 8 \pi}\left|
\sum_i {g_{i\alpha } g_{i\beta } m_{N_i} \over
m_{\delta_1}^2+m_{N_i}^2} \right|^2 \ . \ee As seen from this
formula (and as discussed in detail in \cite{celine}), the
lightest $N_i$ is in general expected to dominate the annihilation
cross section so we obtain \be  {\rm Max}[g_{1 \beta}]\sin \alpha
\sim 5 \times 10^{-4}\left( {m_{N_1} \over {\rm MeV}}\right)^{1/2}
\left( {\langle \sigma v_r \rangle \over 3 \cdot 10^{-26} {\rm
cm}^3{\rm sec}^{-1}}\right)^{1/4} (1+{m_{\delta_1}^2 \over
m_{N_1}^2})^{1/2}\ . \label{ggg}\ee

\begin{table}
\begin{center}
\begin{tabular}{|l|l|l|}
\hline {\rm particle} & {\rm mass} & coupling to
$\bar{N}_i\nu_\alpha$
\\\hline \hline$\delta_1$ & $m_{\delta_1} $ & $-{\sin \alpha \over
\sqrt{2}}g_{i \alpha} $ \\
$\delta_2$ & $m_{\delta_2} $ & ${\cos \alpha \over
\sqrt{2}}g_{i \alpha}$ \\
$\phi_2$ & $m_{\phi_2} $ & ${i \over \sqrt{2}}g_{i \alpha}$\\
\hline
\end{tabular}
\caption{Neutral scalars of the model \label{tab:partcont}.}
\end{center}
\end{table}
As discussed in \cite{celine}, the terms in Eq.~(\ref{sterileL})
 give a Majorana mass to neutrinos through a one loop diagram.
 Using Eq.~(2) of \cite{celine}, we find
\begin{align}  (m_\nu)_{\alpha \beta}=& \sum_i
{g_{i\alpha }g_{i\beta } \over 32 \pi}m_{N_i}\Big[ \sin^2 \alpha
({m_{\delta_2}^2 \over m_{N_i}^2-m_{\delta_2}^2}\log {m_{N_i}^2
\over m_{\delta_2}^2} - {m_{\delta_1}^2 \over
m_{N_i}^2-m_{\delta_1}^2}\log {m_{N_i}^2 \over m_{\delta_1}^2})
\nonumber\\
&+{m_{\phi_2}^2 \over m_{N_i}^2-m_{\phi_2}^2}\log {m_{N_i}^2 \over
m_{\phi_2}^2}-{m_{\delta_2}^2 \over m_{N_i}^2-m_{\delta_2}^2}\log
{m_{N_i}^2 \over m_{\delta_2}^2}\Big]\ .\label{MnUGen}
\end{align}
Notice that  neutrino mass and dark matter annihilation are given
by the same coupling. That is why these two seemingly different
quantities are linked. Such a relation is a feature of the
scenario in \cite{celine} which is embedded within the present
model.

 Taking $m_{\delta_1}^2<m_{N_i}^2\ll
m_{\delta_2}^2$ and using Eq.~(\ref{con2}), we find \be
(m_\nu)_{\alpha \beta}\simeq\sum_i {g_{i\alpha }g_{i\beta } \over
32 \pi}m_{N_i}\Big[ \sin^2 \alpha({m_{\delta_2}^2 \over
m_{N_i}^2-m_{\delta_2}^2}\log {m_{N_i}^2 \over m_{\delta_2}^2} -
{m_{\delta_1}^2 \over m_{N_i}^2-m_{\delta_1}^2}\log {m_{N_i}^2
\over m_{\delta_1}^2}-1)-{\lambda_2}{v_H^2 \over m_{\phi_2}^2}
\Big]\ .\label{mNu}\ee


Three situations are imaginable: (1) $\lambda_2
v_H^2/(m_{\phi_2}^2) \gg \sin^2\alpha \log(m_{\delta_2}^2/m_N^2)$;
(2) $\lambda_2 v_H^2/( m_{\phi_2}^2) \sim \sin^2 \alpha \log
(m_{\delta_2}^2/m_N^2)$; and (3) $\lambda_2 v_H^2/( m_{\phi_2}^2)
\ll \sin^2\alpha \log(m_{\delta_2}^2/m_N^2)$. If we proceed with
case (1) and combine Eq.~(\ref{ggg}) with Eq.~(\ref{mNu}), we find
that the preassumption ({\it i.e.,} $\lambda_2
v_{H}^2/m_{\phi_2}^2 \gg\sin^2\alpha \log(m_N^2/m_{\delta_2}^2)$)
implies $m_{N_1}\ll 1~{\rm MeV}$ which might be at odds with
nucleosynthesis data \cite{raffelt}. Situations (2) and (3) imply
\be m_{N_1} \sim (1 ~{\rm MeV}) \left({3 \cdot 10^{-26}~{\rm
cm}^3~{\rm sec}^{-1} \over \langle \sigma v_r\rangle
}\right)^{1/4}\left({25 \over \log
m_{N_1}^2/m_{\delta_2}^2}\right)^{1/2} \left( {m_\nu \over
\sqrt{\Delta m_{atm}^2}}\right)^{1/2}\left(1+{m_{\delta_1}^2 \over
m_{N_1}^2}\right)^{-1/2}\ee which is the same conclusion reached
in Ref.~\cite{celine}:
$$ m_{\delta_1}<m_{N_1}\sim {\rm few~MeV}\ ,$$
and can be made compatible with the nucleosynthesis data
\cite{serpico}. As discussed in \cite{celine}, in order to
reproduce the neutrino data at least two $N_i$ should be present.
Within minimalistic scenario with only two $N_i$, one of the
active neutrino mass eigenvalues vanishes which means the neutrino
mass scheme is hierarchical. Assuming that annihilation $\delta_1
\delta_1 \to \nu_L \nu_L$ is dominated by $N_1$, the only bound on
the parameters comes from the neutrino mass matrix. Taking
$m_{\delta_2}^2\simeq m_{\phi_2}^2 \simeq m_{\phi^-}^2 \sim
m_{electroweak}^2$, we obtain  \ba \left\{
\begin{matrix} g_{2 \alpha}\sim 2 \times 10^{-4} \left( {m_\nu
\over \sqrt{\Delta m_{atm}^2}}\right)^{1/2} \left({m_{N_2} \over
10~{\rm MeV} }\right)^{- 1/2} {1\over \sin \alpha} ~~~~ & 1~{\rm
MeV}<m_{N_2}\ll m_{\delta_2} \cr  g_{2 \alpha}\sim 7 \times
10^{-6} \left( {m_\nu \over \sqrt{\Delta m_{atm}^2}}\right)^{1/2}
\left({ m_{N_2}\over 100~{\rm GeV}}\right)^{1/2} {1\over \sin
\alpha} ~~~~ &m_{\delta_2} \ll m_{N_2}\ll
m_{\delta_2}\sin\alpha/\lambda_2 \cr g_{2 \alpha}\sim 4 \times
10^{-6} \left( {m_\nu \over \sqrt{\Delta m_{atm}^2}}\right)^{1/2}
\left({ m_{N_2}\over 100~{\rm GeV}}\right)^{-1/2} {1\over
\sqrt{\lambda_2}} ~~~~ &m_{\delta_2}\sin\alpha/\lambda_2 \ll
m_{N_2}
\end{matrix} \right.\label{g2g2g2} \ea

Notice that neither the constraints from the neutrino data nor the
one from the dark matter abundance yields any bound on the masses
of the components of $\Phi$, {\it i.e.,}  $m_{\phi^-}$,
$m_{\phi_1}$ or $m_{\phi_2}$. In principle, as long as $\sin
\alpha \sim v_H m_{\eta \phi}/(m_{\phi_1}^2-m_\eta^2)$ remains
smaller than few$\times 10^{-4}$ (see Eq.~(\ref{ggg})), the
components of $\Phi$ can be arbitrarily heavy. Notice however that
for $ m^2_{\phi_1}/m_{Electroweak}^2 \to \infty$, $m_{\eta \phi}$
also grows as $\sin \alpha m_{\phi_1}^2/v_H$. Moreover for
$m_{\phi_1}^2 \gg m_{electroweak}^2$, from Eq.~(\ref{2ALPHA}) we
observe that in order to maintain $m_{\delta_1}^2$ around $(1~{\rm
MeV})^2$, a fine tuned cancelation between $m_\eta^2$ and
$(m_{\eta \Phi} v_H)^2/(m_{\phi_1}^2-m_\eta^2)\sim \sin^2 \alpha
(m_{\phi_1}^2-m_\eta^2)$ is required. The degree of fine tuning is
$m_{\delta_1}^2/(\sin^2 \alpha m_{\phi_1}^2)\simeq 10^{-4} (1~{\rm
TeV})^2/m_{\phi_1}^2$ so to keep the degree of fine tuning
agreeable, the range $m_{\phi_1}^2\ll ({\rm TeV})^2$ is more
desirable.

After replacing $g \to g \sin \alpha/\sqrt{2}$, all discussion in
Ref.~\cite{celine} on the falsifiability of the model via low
energy experiments can be repeated here, too. In addition to the
phenomenological implications discussed in \cite{celine},  the
heavy states  will cause a number of potentially observable
effects which we discuss in the next section.


\section{Phenomenological implications \label{Impli}}
In Ref.~\cite{celine,celine1}, a number of phenomenological
consequences of the low energy content of the scenario have been
discussed. The same consideration applies when the scenario is
embedded within a model.

Large scale structure arguments imply that the dark matter
candidate has to be  heavier than few keV \cite{LSS,celine}.
Although our dark matter candidate can affect the supernova
evolution, it can be accommodated within current uncertainties
\cite{celine}. Detection of future supernova explosions will be a
powerful probe for this scenario. Finally, since there are new
light particles $N_i$ and $\delta_1$ which couple to ordinary
neutrinos, they can show up in light meson decays as missing
energy signal ($K(\pi) \to \ell N_i \delta_1$). Present bounds on
the coupling \cite{KPI} are too weak to rule out our model
\cite{celine}. Eventually by  improving the bound on $K\to \ell
+{\rm missing~energy}$ and $\pi \to \ell +{\rm missing~energy}$,
the condition in (\ref{ggg}) can be tested \cite{yasaman}. Thus, a
careful analysis of the KLOE results is imperative for testing
this model.

From Eq.~(\ref{mixing-ETA-PHI1}), we observe that while real
singlet, $\eta$, and the CP-even component of $\Phi$, $\phi_1$
mix, the CP-odd component, $\phi_2$ is decoupled and does not
enter the mixing matrix. This is expected because we took the
Lagrangian to be CP-even. Had we allowed $m_{\eta \phi}$ and
$\lambda_2$ to be complex, a mixing between $\eta$ and $\phi_2$
would also appear. The coupling of $\phi^0=(\phi_1+i
\phi_2)/\sqrt{2}$ to $Z$ boson is of the following form:
$$ (\phi_2 \partial_\mu \phi_1 -\phi_1 \partial_\mu \phi_2)Z^\mu\ .$$
The absence of a component
 of $\phi_2$ in $\delta_1$ has several consequences:
 1) Since
$\phi_2$ is heavy, no new decay mode for the $Z$ boson appears. 2)
The interaction of dark matter with electrons and nuclei does not
take place at tree level, making direct detection as well as
trapping in dense environments such as the Sun more challenging.
Moreover, the mass of the dark matter candidate in this model is
too small to lead to a detectable recoil energy in direct dark
matter search experiments. Through couplings to the Higgs boson
({\it i.e.,} $\lambda_i$), the dark matter can interact with
nuclei.  If the couplings are relatively large, dark matter can be
trapped inside the Sun. Relatively high abundance of dark matter
in the halo can give rise to a neutrino flux of few MeV
potentially detectable at Super-Kamiokande. The corresponding
bound has been studied in \cite{sergio} but the present bound is
too weak to probe $\langle \sigma v_r\rangle\sim 3\cdot 10^{-26}
{\rm cm}^3 {\rm sec}^{-1}$.

 In addition to the effects caused by light particles of the
model ({\it i.e.,} $N_i$ and $\delta_1$), due to the presence of
the heavier particles within the present model, a number of new
effects will appear which we discuss below.

 \subsection{ Annihilation into electron positron pair}
 In the present model, the annihilation of dark matter
  to the $e^- e^+$ pair can take
 place only at the loop level.
There are two one-loop diagrams contributing to this annihilation.
In one of these two, the vertices involving $\delta_1$ are of the
form $-(g_{ie}\sin \alpha /\sqrt{2}) \delta_1 \bar{N}_i \nu_e$ and
in the other, only the gauge couplings enter.
 Depending on the range of parameters, either
of the following contributions to the annihilation cross section
can be dominant:
$$\langle \sigma v_r \rangle\sim
\left({e^4 \over 16 \pi^2\sin^4\theta_W}\right)^2 {(m_{\delta_1}
v_r)^2 \over 8\pi m_W^4}\sin^4 \alpha\ \ \ \  {\rm or}\ \ \  \sim
\left({e^2 \sum_i |g_{ i e}|^2\over 16
\pi^2\sin^2\theta_W}\right)^2 {(m_{\delta_1} v_r)^2 \over 8\pi
m_W^4}\sin^4 \alpha .$$   Since SLIMs are scalars and the
annihilation into an electron positron pair does not flip the
chirality, the annihilation is a $p$-wave process so the cross
section is proportional to the velocity square ($v_r^2$).
 In order to explain the 511 keV line from the center of the
 galaxy the ratio  $\sigma({\rm Dark ~Matter+Dark ~Matter }\to
e^-+ e^+)/ \sigma_{tot}$ has to be of order of $10^{-4}$. The loop
suppression factor [{\it i.e.,} $e^4/(16 \pi^2\sin^2 \theta_w)^4$]
is smaller than this amount so the 511 keV line cannot be
explained within this model. Moreover, because of the low
annihilation cross section, we do not expect any detectable photon
radiation from the emitted $e^-e^+$ pairs. Notice that the
annihilation of non-relativistic SLIMs into $\mu^+\mu^-$ cannot
take place because of the low mass of the particles.

\subsection{Annihilation into photon pair}
{There are several one-loop diagrams that contribute to the
annihilation into a photon pair. Each diagram is divergent but the
divergences cancel each other and at the end of the day, the cross
section can be written as \be \sigma(\delta_1\delta_1 \to \gamma
\gamma)\sim{e^8 \sin^4 \alpha \over 8 \pi (16 \pi^2)^2\cos^4
\theta_W}{m_{\delta_1}^2 \over m_W^4}\sim{\rm few} \times 10^{-41}
\left( {M_{\delta_1}\over {\rm MeV}}\right)^2 \sin^4 \alpha~{\rm
cm}^3/{\rm sec} \ . \ee  The value of annihilation cross section
is too small for the Fermi telescope (formerly known as GLAST) to
detect  a signal for annihilation into a photon pair (see, {\it
e.g.,} Fig 4 of \cite{stefano}).}

\subsection{Lepton Flavor Violating rare decay}

In both of the models presented in this paper, there is a charged
scalar that couples to the charged leptons through $g_{i \alpha}
\bar{N}_i \ell_{L\alpha}\phi^-$. This coupling leads to the Lepton
Flavor Violating rare decays, $\mu \to e\gamma$, $\tau \to \mu
\gamma$ and $\tau \to e \gamma$. Using the formulas in
\cite{Lavoura}, we find\be \Gamma(\ell_\alpha \to \ell_\beta
\gamma)={m_\alpha^3 \over 16\pi}|\sigma_R|^2\ , \ee where
$$\sigma_R=\sum_i g_{i \alpha } g_{i\beta }^* {iem_\alpha \over 16
\pi^2 m_{\phi^-}^2} K(t_i)\ , $$ where
$t_i=m_{N_i}^2/m_{\phi^-}^2$ and \be K(t_i)={2 t_i^2+5t_i-1 \over
12 (t_i-1)^3}-{t_i^2 \log t_i \over 2(t_i-1)^4} \ .\label{k(t)}
\ee The following three limits are of interest: (1) $t_i\to 0$ for
which $K(t_i)\to 1/12$; (2) $t_i \to \infty $  which implies
$K(t_i)\to 1/(6 t_i)$ and (3) $t_i \sim 1$ which implies
$K(t_i)\sim 1$.

As discussed earlier, the mass of at least one of $N_i$ has to be
around 1~MeV. The mass of the other $N_i$ can be larger. As long
as $m_{N_2}^2\ll m_{\phi^-}^2$, we can write
\begin{align}
{\rm Br}(\mu \to e \gamma)&\sim 2 \times 10^{-4}|\sum_i g_{\mu
i}g_{e i}^*|^2 \left({100~{\rm GeV} \over m_{\phi^-}}\right)^4 \\
{\rm Br}(\tau \to \ell_\alpha \gamma)&\sim 5 \times 10^{-5}|\sum_i
g_{i\tau }g_{i\alpha }^*|^2 \left({100~{\rm GeV} \over
m_{\phi^-}}\right)^4.
\end{align}
From PDG booklet \cite{pdg}, we read
\begin{align}
{\rm Br}(\mu \to e \gamma)&<1.2\times 10^{-11} \\
{\rm Br}(\tau \to e \gamma)&< 1.1 \times 10^{-7} \\
{\rm Br}(\tau \to \mu \gamma)&<6.8 \times 10^{-8}\ .
\end{align}
From Eqs.~(\ref{ggg},\ref{g2g2g2}), we realize that the present
bounds can be readily satisfied. A particularly interesting range
 is the following: \be m_{\phi^-}\sim 100~{\rm GeV} \ \ \
g_{i \mu},g_{i\tau}\sim{\rm few}\times 10^{-2} \  \ {\rm and} \ \
g_{i e}\sim{\rm few}\times 10^{-3},\label{gParticular}\ee which
leads to the LFV rates detectable in near future \cite{MEG}.
Notice that a little hierarchy of $g_{i e}\sim (\Delta
 m_{sol}^2/\Delta m_{atm}^2)^{1/2} g_{i \mu}\sim 0.1 g_{i \mu}$
 is desirable for normal hierarchical mass scheme.

 In future if these LFV rare decays are discovered, it will be
 possible to derive information on the flavor structure of the
 couplings. By combining the information from the flavor
 structure of the neutrino mass matrix
 with the information from LFV searches the models can be  probed
\cite{yasaman}.

\subsection{Magnetic dipole moment of the muon}
In the models presented in this paper, the magnetic dipole moment
of the muon obtains a contribution via coupling $g_{i \mu }
\bar{N}_i \mu (\phi^-)^\dagger$:
$$ \delta{g-2\over 2}=\sum_i {|g_{i\mu }|^2 \over 16
\pi^2}{m_\mu^2 \over m_{\phi^-}^2}K(t_i)\ ,$$ where $K(t_i)$ is
defined in Eq.~(\ref{k(t)}).
 For $t_i \ll 1$,
which is what we expect for our model, we can write
$$\delta{g-2\over 2}=5\times 10^{-12}{\sum_i |g_{i \mu }|^2 \over
10^{-2}}\left({100~{\rm GeV} \over m_{\phi^-}^2}\right)^2$$ so the
present bounds can be readily satisfied. The predicted value is
below the present bound by two orders of magnitude.

\subsection{Dark matter self interaction}
The $\lambda'_i$ couplings can give rise to self interaction of
the SLIM particles:
$$ \langle \sigma(\delta_1\delta_1\to \delta_1\delta_1) v\rangle
\sim {\rm Max}[{|\lambda'_1|^2 \sin^4 \alpha\over 8 \pi
m_{\delta_1}^2},{|\lambda'_2|^2 \cos^4 \alpha\over 8 \pi
m_{\delta_1}^2},{|\lambda'_3|^2 \sin^2\alpha \cos^2 \alpha \over 8
\pi m_{\delta_1}^2}]\ .$$ From considering the merging of the
galaxy clusters, bounds on $\sigma/m_{DM}$ have been derived
\cite{bullet}: $\sigma/m_{DM}\stackrel{<}{\sim}1~{\rm cm^2/g}$.
The bound implies
$$|\lambda_1'|^2 \sin^4\alpha,|\lambda'_2|^2\cos^4 \alpha,|\lambda'_3|^2
\sin^2\alpha \cos^2 \alpha\stackrel{<}{\sim} 10^{-4}\ .$$ There
have been some suggestions to employ self interaction of dark
matter with $\sigma/m_{DM}=(0.5-5)~ {\rm cm^2/g}$ to explain the
observed mass profiles of the galaxies \cite{alleviate}. The
possibility can be readily accommodated within the present model.
\subsection{Production at LHC}
Last but not least, the components of doublet $\Phi$  can  in
principle be produced at the LHC through electroweak interactions
in pairs. All the new particles will eventually decay into SLIM
which appears as missing energy signal at the detector.

Remember that  at least one of the right-handed neutrinos has to
be lighter than the doublet. As a result, $\phi^-$ can decay into
$\ell_\beta^-$ and $N_i$ \be \Gamma(\phi^- \to \ell_\beta^-
N_i)\propto |g_{i \beta}|^2 \ .\ee Thus, by studying the two body
decay mode into a charged lepton, the couplings to light
right-handed neutrinos can be determined. The same couplings enter
the neutrino mass matrix and the formulas for the rates of the LFV
rare decays so the model can be cross checked by studying the
correlations \cite{yasaman}. The light $N_i$ will eventually decay
into a SLIM and ordinary neutrinos: $N_i \to \nu \delta_1$. Both
of the decay products escape detection and  appear as missing
energy signal. As discussed in the end of sect.~\ref{RealSLIM},
there is no upper bound on the masses of the components of the
doublet but from the fine tuning considerations, mass scale of 100
GeV is more desirable. Thus, if the LHC does not find any signal
for such a doublet, the model cannot be falsified however, it will
require a higher degree of fine tuning.

Within this scenario Higgs can have invisible decay modes $H \to
\delta_1 \delta_1$. If ${\rm Max}[ \cos^2 \alpha\lambda_3,
\sin^2\alpha \lambda_{i ~ {\rm with}~ i\ne 3} ] \stackrel{>}{\sim}
m_b/v_H\simeq 0.02$, this decay mode can dominate over the
standard decay mode $H\to b \bar{b}$.

\section{Conclusions \label{con}}
We have proposed a model that encompasses the scenario introduced
in \cite{celine} for linking dark matter problem with the neutrino
mass puzzle.  The model inherits the features of the scenario: It
contains a scalar (SLIM) and one (or more) right-handed
neutrino(s) with masses lower than $O(1~{\rm MeV})$. As in
Ref.~\cite{celine}, combining data on neutrino mass scale with
dark matter abundance puts a lower bound on the couplings between
the right-handed neutrinos with the SLIM which makes the model
testable by low energy experiments. Since in the model the mass of
the dark matter is much smaller than the nucleon mass, we expect
null result in direct searches for the dark matter which are based
on measuring the recoil energy of scattering of dark matter
particles off nuclei in a sample.

The heavy particles of the model which consist of the components
of a scalar doublet can  in principle be produced at the
accelerators. The charged component of the doublet, $\phi^-$,
which couples to  charged leptons via the same LFV couplings that
enter the neutrino mass matrix, can give rise to LFV processes,
$\ell_i\to \ell_j \gamma$ at one loop level. Studying the decay
modes of $\phi^-$, the flavor structure of its Yukawa coupling can
be derived and can be tested against the information obtained from
flavor structure of the neutrino mass matrix. We have not derived
any upper bound on the masses of the doublet so if the LHC reports
a null result, the model will not be ruled out. We have however
shown that the fine tuning arguments points towards low mass scale
doublets and therefore towards a rich phenomenology at the LHC.

\begin{acknowledgments}
I would like to thank S. Pascoli, T. Hambye, M. Schmidt and  M. M.
Sheikh-Jabbari for useful discussions.
\end{acknowledgments}



\begin{thebibliography}{9}
\bibitem{celine}
  C.~Boehm, Y.~Farzan, T.~Hambye, S.~Palomares-Ruiz and S.~Pascoli,
  Phys.\ Rev.\  D {\bf 77} (2008) 043516.
  \bibitem{linkingModels}
L.~M.~Krauss, S.~Nasri and M.~Trodden,
  Phys.\ Rev.\  D {\bf 67} (2003) 085002
  [arXiv:hep-ph/0210389];
K.~Cheung and O.~Seto,
  Phys.\ Rev.\  D {\bf 69} (2004) 113009
  [arXiv:hep-ph/0403003];
T.~Asaka, S.~Blanchet and M.~Shaposhnikov,
  Phys.\ Lett.\  B {\bf 631} (2005) 151
  [arXiv:hep-ph/0503065];
E.~J.~Chun and H.~B.~Kim,
  JHEP {\bf 0610} (2006) 082
  [arXiv:hep-ph/0607076];
J.~Kubo and D.~Suematsu,
  Phys.\ Lett.\  B {\bf 643} (2006) 336
  [arXiv:hep-ph/0610006];
J.~Kubo, E.~Ma and D.~Suematsu,
  Phys.\ Lett.\  B {\bf 642} (2006) 18
  [arXiv:hep-ph/0604114];
T.~Hambye, K.~Kannike, E.~Ma and M.~Raidal,
  Phys.\ Rev.\  D {\bf 75} (2007) 095003
  [arXiv:hep-ph/0609228];
N.~Sahu and U.~Sarkar,
 Phys.\ Rev.\  D {\bf 78}, 115013 (2008)
 [arXiv:0804.2072 [hep-ph]].


\bibitem{ourselves}
Y. Farzan, S. Pascoli and M. Schmidt, 
  arXiv:1005.5323 [hep-ph]; to appear in JHEP.


\bibitem{Ma}
 E.~Ma,
  Phys.\ Rev.\  D {\bf 73} (2006) 077301
  [arXiv:hep-ph/0601225].
\bibitem{G-H}
H.~E.~Haber,
  Phys.\ Rev.\  D {\bf 67} (2003) 075019
  [arXiv:hep-ph/0207010].



\bibitem{raffelt}
  S.~Hannestad and G.~G.~Raffelt,
  JCAP {\bf 0611} (2006) 016
  [arXiv:astro-ph/0607101].
\bibitem{serpico}
  P.~D.~Serpico and G.~G.~Raffelt,
  Phys.\ Rev.\  D {\bf 70} (2004) 043526
  [arXiv:astro-ph/0403417].

\bibitem{celine1}
C.~Boehm and R.~Schaeffer,
  arXiv:astro-ph/0410591;
C.~Boehm, P.~Fayet and R.~Schaeffer,
  Phys.\ Lett.\  B {\bf 518} (2001) 8
  [arXiv:astro-ph/0012504].
\bibitem{LSS}
U.~Seljak, A.~Makarov, P.~McDonald and H.~Trac,
  Phys.\ Rev.\ Lett.\  {\bf 97} (2006) 191303
  [arXiv:astro-ph/0602430];
A.~Boyarsky, J.~Lesgourgues, O.~Ruchayskiy and M.~Viel,
  JCAP {\bf 0905} (2009) 012
  [arXiv:0812.0010 [astro-ph]].

\bibitem{KPI}
D.~I.~Britton {\it et al.},
  Phys.\ Rev.\  D {\bf 49} (1994) 28;
V.~D.~Barger, W.~Y.~Keung and S.~Pakvasa,
  Phys.\ Rev.\  D {\bf 25} (1982) 907;
 G.~B.~Gelmini, S.~Nussinov and M.~Roncadelli,
  Nucl.\ Phys.\  B {\bf 209} (1982) 157.


\bibitem{yasaman}
Y.~Farzan,
  arXiv:1009.1234 [hep-ph].


\bibitem{sergio}
S.~Palomares-Ruiz and S.~Pascoli,
  Phys.\ Rev.\  D {\bf 77} (2008) 025025
  [arXiv:0710.5420 [astro-ph]].

\bibitem{stefano}
S.~Profumo,
  Phys.\ Rev.\  D {\bf 78}, 023507 (2008)
  [arXiv:0806.2150 [hep-ph]].

\bibitem{Lavoura}
 L.~Lavoura,
  Eur.\ Phys.\ J.\  C {\bf 29} (2003) 191
  [arXiv:hep-ph/0302221].
\bibitem{pdg}
  C.~Amsler {\it et al.}  [Particle Data Group],
  Phys.\ Lett.\  B {\bf 667} (2008) 1.



\bibitem{MEG}
http://meg.web.psi.ch/index.html; http://superb.kek.jp/;
  M.~A.~Giorgi,
  J.\ Phys.\ Conf.\ Ser.\  {\bf 171} (2009) 012022.



\bibitem{bullet}
 S.~W.~Randall, M.~Markevitch, D.~Clowe, A.~H.~Gonzalez and M.~Bradac,
  arXiv:0704.0261 [astro-ph].
\bibitem{alleviate}
 R.~Dave, D.~N.~Spergel, P.~J.~Steinhardt and B.~D.~Wandelt,
  Astrophys.\ J.\  {\bf 547} (2001) 574
  [arXiv:astro-ph/0006218];










\end{thebibliography}
\end{document}